\documentclass[journal]{IEEEtran}
\usepackage{amssymb,amsmath,amsthm}
\usepackage{graphicx}
\usepackage{pb-diagram}
\usepackage{color}
\usepackage{multirow}
\usepackage{ulem}

\theoremstyle{plain}

\theoremstyle{definition}

\begin{document}
\title{Subthermal switching with nanomechanical relays}
\author{Dincer Unluer and  Avik W. Ghosh
\thanks{All authors are with the Dept. of Electrical and Computer Engineering, University of Virginia, Charlottesville, VA 22904.}
\thanks{This work has been supported by NanoStar.}}

\maketitle
\begin{abstract}
We present a physical model for electronic switching in cantilever based nano-electro-mechanical field effect transistors, focusing on the steepness of its switching curve. We find that the subthreshold swing of the voltage transfer characteristic is governed by two separate considerations - the ability of the charges to correlate together through dipolar interactions and amplify the active torque, versus the active pull-in forces that drive an abrupt phase transition and close the air gap between the tip of the cantilever and the drain. For small sized relays, dipolar and  short-range Van Der Waals `sticking' forces dominate, while for longer cantilevers the capacitive energy acquires a major role. The individual pull-in and pull-out phases demonstrate a remarkably low subthreshold swing driven by the capacitive forces, sharpened further by dipolar correlation. The sharp switching, however, comes at the expense of strong hysteresis as the metastable and stable states interchange along the forward and reverse phases of the voltage scan.  
\end{abstract}


\section{Introduction}
\IEEEPARstart{D}{igital} electronics is fundamentally constrained by the cost of binary switching $Q\Delta V$. The number of charges $N = Q/q$ is quite large (between 10-100 in the past 10 years) for a n-MOS transistor, and ramps up rapidly to a few thousand for a fan-out 4 device with interconnect, because of the need to charge up the entire interconnect capacitance and drive an analogous transistor downstream \cite{1250,4215159}. $\Delta V$ on the other hand is constrained by the Boltzmann limit, and corresponds to a switching slope of $k_BT\ln{10}/q \sim 60$ mV/decade. Two paths to bypassing these constraints while maintaining an acceptable reliability are (i) {\it{correlated}} switches as in nano-magnetic logic \cite{Salahuddin2013}, where several units, in this case spins, can offer solidarity against thermal fluctuations, and yet limit the energy cost by correlating together through their internal coupling fields; and (ii) {\it{phase-transition}} switches that operate either far from equilibrium or near a phase transition point, so that a small voltage precipitates a large change, such as through an internal voltage amplification.

In order to lower the power supply voltage $V_{DD}$ and thereby cut static and dynamic power dissipation, we ultimately need devices with lower sub-threshold swings and steeper gate voltage transfer curves beyond what the Boltzmann limit posits. This has led to the exploration of various unconventional switching devices such as tunnel field effect transistors \cite{reddick494}, impact ionization devices \cite{1372710}, ferroelectric transition switches \cite{4796789} and chiral tunneling devices based on graphene pn-junctions \cite{sajjad123101}, all of which purport to achieve swings lower than the Boltzmann limit. In this regard, mechanical switches or relays are noteworthy, as they can swing between near zero off-currents and high on-currents at very low $V_{DD}$ values \cite{5575378}. Such nano-electro-mechanical field effect transistors (NEMFETs) are relevant as low power switches \cite{5424218}, especially when integrated into hybrid CMOS circuits to minimize leakage current from the power supply supply in the OFF state \cite{6571905}. Indeed, ultra-low subthreshold swings have been demonstrated using suspended-gates \cite{1609384}, and micro-relays \cite{5424218,5361249},  along with complimentary relays for logic applications with larger voltage margins and reliability \cite{5419997}. There now are multiple designs including the printable relay \cite{6131636} and the curved relay by IBM \cite{6411561}. Relays are also known to operate as active elements in turning on voltage gated ion channels in axonal systems. In addition, it is noteworthy that the voltage gating of sodium channels involving a mechanical relay was measured many decades ago at a subthreshold swing of $20$ mV/dec \cite{Hodgkin1952currents,Doyle1998,MacKinnon2003}.

The action of a NEMFET is significantly different from a conventional electrostatically driven field effect transistor, and needs new physics-based predictive models. A number of empirical models have already been proposed based on 1D electrostatics \cite{5361249,4418930}, nonlinear oscillator equations \cite{Hibert:2002:MDL:846240.850270,1406573,1609380}, pull-in instability \cite{0960-1317-16-2-025,4443702}, deformation characteristics of microstructures under electrostatic loads \cite{1007404}, and techniques from chaos and bifurcation theory in finite-elements \cite{luo:77,alurutalk,Li:2001:0924-4247:278}. Our purpose in this paper is to explain the electron transport physics in a relay switch, starting with the microscopic Landauer equation that also elegantly explains the limits on the subthreshold swing. We argue that a dipole coupled, cantilever based relay can function as both a correlated switch and a phase transition switch.

\section{Device Structure and transport equations}
Nanoelectromechnanical switches exist in various flavors - {\it{movable gates}} or {\it{movable channels}}. Movable gate switches are driven by the capacitive energy $CV_G^2/2$, where the capacitance $C =A\Bigl(t_{ox}/\epsilon_{ox} + t_{air}/\epsilon_0\Bigr)^{-1}$ is divided serially between a fixed oxide and a variable air gap. Upon application of a gate bias, a force $F = -\partial U/\partial V_G$ alters the air gap from its equilibrium value $t_0$, counterbalanced by a spring which exerts a restoring force $-k(t_{air}-t_0)$ \cite{rebeiz2004rf}. At a critical distance where the electrostatic pull-in overcomes the restoring force, there is an abrupt closing of the air gap, precipitating a sudden rise in capacitance. Since the pull-in acts on a capacitive structure, one can play with material properties of the gate oxide and shape its potential landscape in a creative way in order to liberate the movable gate easily \cite{6656832}. Since the channel itself is fixed, its OFF current is limited by electrostatic depletion, and for a small channel suffers from the usual source-to-drain tunneling, although the removal of the movable gate reduces the gate oxide leakage significantly.  

In a movable channel transistor, the idea is to physically separate the channel away from the drain so that in addition to electrostatic depletion, the OFF current can be reduced substantially with an added quantum mechanical tunneling barrier. In fact, it is straightforward to show that their electrostatic and conformational modes and thus their subthreshold swings act in parallel, $S^{-1} \approx S^{-1}_{el} + S^{-1}_{conf}$ \cite{nl035109u}, assuming a clear separation of electronic and conformational time constants. We have shown in the past \cite{nl035109u} that the conformational mode, besides aiding the electrostatic turn-off, allows an intrinsically high transconductance, and thus a low subthreshold swing
\begin{equation}\
S_{conf} = \displaystyle\Biggl(\frac{k_BT}{q}\ln{10}\Biggr)\Biggl\langle\frac{q(t_{air}(\theta) + t_{ox})}{\mu C(\theta)}\Biggr\rangle
\end{equation}
where $C$ is a correlation function that we will describe later. The cantilever FET can function as both a correlated switch and a phase transition switch. The presence of the dipole $\mu$ means that the subthreshold swing can be reduced by a factor $qt_{ox}/\mu \propto t_{ox}/L$ for longer cantilevers, exploiting the property that many charges are physically moving with the cantilever for the price of one. This is, in fact, the suggested mechanism behind the observed sharp switching in voltage-gated sodium channels \cite{MacKinnon2003}. A separate mechanism not analyzed in \cite{nl035109u} arises from an additional `pull-in' that sucks in the end of the cantilever once it is physically proximal with the drain. The short-ranged Van der Waals force and the longer ranged capacitive forces cause an abrupt shrinkage of the air gap and an exponential rise in tunneling current, reducing the subthreshold swing further by a factor proportional to $dt_{air}/d\theta$, which vanishes near the contact point, giving a near vanishing subthreshold swing. 

The Van der Waals pull-in helps subvert thermal fluctuations by stabilizing the attractive forces. Resetting the device however, requires circumventing adhesion, which creates a sizable hysteresis loop. The switch thus has extremely sharp transitions along the walls of the hysteresis curve, but its loop-averaged subthreshold swing is quite large and problematic. Fixing this requires creative material and geometry design. Our purpose in this paper is to develop the mathematical model and the physical picture behind the entire cantilever dynamics in its ON, OFF and transitory states, the geometry of the hysteresis and its dependence on material parameters as well as the voltage scan rate.

\begin{figure}[t]
\centering
\includegraphics[width=\linewidth]{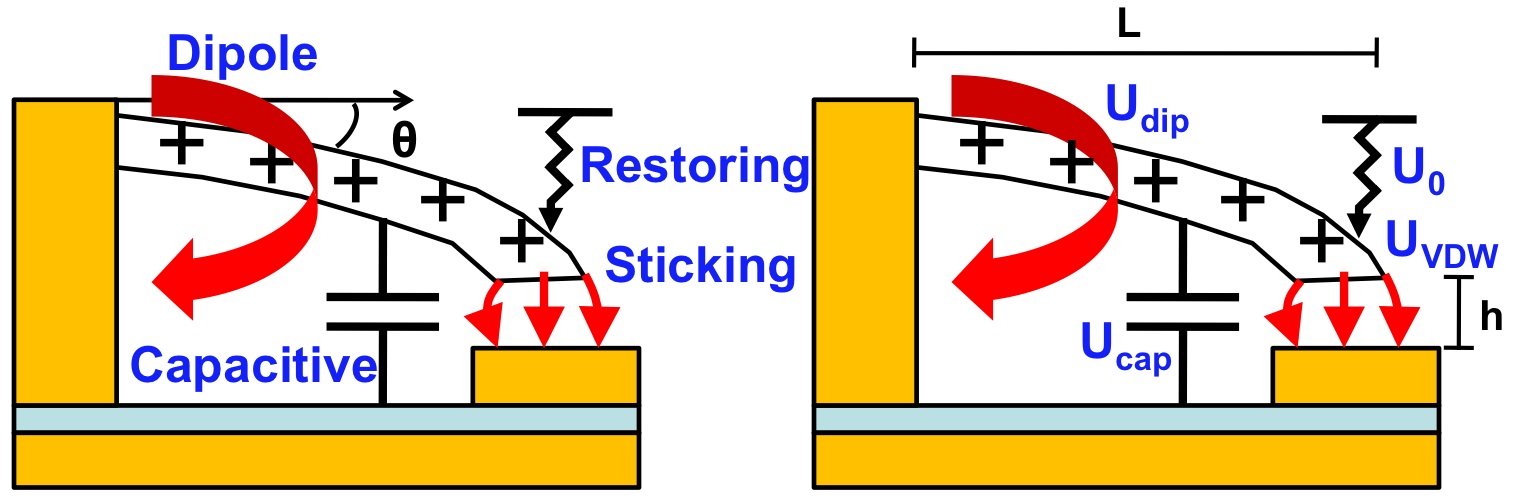}
\caption{(Left) Schematic picture of cantilever showing all the forces operating on it. (Right) Associated energy terms, spelled out in Eq.~\ref{pots}}
\label{fig:geometry}
\end{figure}

The electron current is determined by the degree of quantum mechanical tunneling across the gap between the cantilever and the drain. We can describe that using the Landauer equation \cite{datta1997electronic}, written in an equivalent but less familiar \cite{BagwellOrlando} form as a convolution between the {\it{zero-temperature}} current and a thermal broadening function $F_T(E) = -\partial f(E)/\partial E$
\begin{equation}
I = \frac{2q}{h}\int_{\mu_1}^{\mu_2} dE M(E) \biggl\langle \displaystyle\bar{T}(E,V_G)\biggr\rangle \otimes F_T(E,V_G)
\label{landauer}
\end{equation}
where $M(E)$ is the mode count in the cantilever, $f$ is the equilibrium Fermi-Dirac distribution, with the $V_G$ dependence in $F_T$ arising from the location of the Fermi energy $E_F$ which shifts with $V_G$ scaled by a gate transfer factor $\alpha_G$. $\mu_{1,2}$ are the bias separated electrochemical potentials in the source and drain contacts, and $\bar{T}(E,V_G)$ is the mode and conformation averaged transmission function controlled by the gate voltage-dependent orientation. $\langle\ldots\rangle$ represents the thermodynamic average of the cantilever orientation $\theta$ over various angles, weighted by its probability distribution $P$ 
\begin{equation}
\biggl\langle \bar{T}(E,V_G)\biggr\rangle = \frac{\displaystyle\int d\theta \bar{T}(E,\theta) P(\theta,V_G)}{\int d\theta P(\theta,V_G)}
\end{equation}
If we assume the mechanical modes in the cantilever are at equilibrium, (i.e., the potential is ramped in time much slower than the relaxation scales in the lattice), $P$ is simply the Boltzmann solution $\propto \exp{[-U(\theta,V_G)/k_BT]}$, even though the electrons flowing are clearly in non-equilibrium.  In general, the cantilever's angular distribution satisfies a thermodynamic equation that needs to be derived properly from a stochastic Lang\'{e}vin equation in the presence of a fast ramp speed of the potential. A flexible cantilever clamped on one end and free on the other satisfies a fourth degree partial differential equation of motion. For an over-damped, rigid cantilever with a single defining angle $\theta$, the equation can be simplified down to second degree and consequently $P$ can be shown to satisfy the Fokker-Planck equation
\begin{equation}
\gamma\frac{\partial P}{\partial t} = \frac{\partial}{\partial \theta}\Biggl(P\frac{\partial U(\theta,V_G)}{\partial \theta}\Biggr)
+ k_BT\frac{\partial^2P}{\partial \theta^2}
\label{eq4}
\end{equation}
where $\gamma$ is the angular damping constant with units J-s. We assume here that all phonons emitted in the cantilever are promptly reabsorbed by a conductive thermal coating coupled to the source, so we can ignore any self-heating and phonon bottleneck effects in the cantilever itself.

The above equations carry all the necessary ingredients for our NEMFET operation. The normal subthreshold swing of $k_BT\ln{10}/\alpha_Gq$ arises from the gate dependent shift in Fermi energy relative to the mode placement $M$ for a fixed orientation, when $U$ is a delta function around a given $\theta$, and $P$ is simply equilibrium Boltzmann. Further reduction in subthreshold swing will arise from the added gate dependence of the conformational modes, hidden in $U(\theta, V_G)$. The hysteresis will arise from a transition out of metastable states with a non-equilibrium distribution $P$ in presence of a time-dependent sweep in gate voltage, $V_G(t)$. Finally, a negative capacitance for a ferroelectric oxide \cite{6656832} will be captured by the gate voltage dependence of the oxide dielectric constant, $\epsilon_{ox}(V_G)$ sitting in $U$, which can introduce its own dynamics near the ferroelectric phase transition point. 

\begin{table}[b!]
\renewcommand{\arraystretch}{1.3}
\caption{Simulation Parameters for Model}
\label{table1}
\centering
\begin{tabular}{| c | c | c | c |}
\hline
$L$ & $5 ~nm$ & $k_BT$ & $0.025 ~eV$ \\
$h$ & $2.5 ~nm$ & $U_{0}^{Bend}$ & $~2$ \\
$W$ & $0.5 ~nm$ & $\vec{\mu}$ & $0.15 ~e\AA$ \\
$t_{air}$ & $1.5 ~nm$ & $\epsilon_{Relay}$ & $3.98 ~eV$ \\
$\epsilon_{air}$ & $1$ & $\epsilon_{Metal Plate}$ & $0.67 ~eV$ \\
$t_{ox}$ & $1 ~nm$ & $\sigma_{Relay}$ & $3.47 ~\AA$ \\
$\epsilon_{ox}$ & $3.9$ & $\sigma_{Metal Plate}$ & $3.144 ~\AA$ \\
\hline
\end{tabular}
\end{table}

\section{Operation of the relay}
The physics of the relay is captured by its transmission and its conformational potential profile. The transmission is limited by electron tunneling into the drain from the edge of the cantilever through the vacuum barrier of width $t_{air}(\theta) = h - L\sin{\theta}$, $h$ being the bare height above the drain before bending and $L$ being the length of the cantilever (Fig. \ref{fig:geometry}). The transmission, using a WKB approximation, is $\bar{T}(E,\theta) \propto exp[-2\kappa t_{air}(\theta)]$. The main physics of the NEMFET dynamics is captured by its potential $U(\theta, V_G)$, which can be separated into a flexural potential $U_0$ representing the tendency of the cantilever to snap back to its horizontal orientation, a gate driven field $\vec{\cal{E}}$ which couples with the dipole moment $\vec{\mu}$ along with capacitive coupling with the back gate to bend the cantilever, and finally a Van Der Waals `pull-in' potential $U_{VDW}$ that causes the cantilever to `stick' to the drain when close enough (Fig. \ref{fig:geometry})
\begin{eqnarray}
U[\theta,V_G] &=& U_0(\theta) + U_{VDW}(\theta) \nonumber\\
& + & U_{dip}[\theta,V_G] + U_{cap}[\theta,V_G] \nonumber\\
U_0(\theta) &\approx& \frac{1}{2}U_{0}^{Bend}\theta^2 \nonumber\\
U_{VDW}(\theta) &=& 4\epsilon\left[\Biggl(\frac{\sigma}{t_{air}(\theta)}\Biggr)^{12} -  \Biggl(\frac{\sigma}{t_{air}(\theta)}\Biggr)^{6}\right] \nonumber\\
U_{dip}[\theta,V_G] &=& -\vec{\mu}\cdot\vec{\cal{E}} = -\mu\Biggl(\frac{V_G\sin\theta}{t_{ox}/\epsilon_{ox} + t_{air}(\theta)/\epsilon_{air}}\Biggr)\nonumber\\
U_{cap}[\theta,V_G] &=& -\frac{1}{2}\Biggl(\frac{A}{t_{ox}/\epsilon_{ox} + t_{air}(\theta)/\epsilon_{air}}\Biggr)V_G^2
\label{pots}
\end{eqnarray}
We assume a simple parabolic potential for the bending dynamics to focus on the basic physics. Higher nonlinear terms would capture the full import of the flexural modes \cite{1097796}, although they do not  provide us with immediate insights \cite{4796329}. It is worth emphasizing that the dipolar torque has a distinct advantage over capacitive forces, as it exploits the vector directionality of the gate field. Even if the oxide is thick enough to normally produce short-channel effects, the gate field is orthogonal to the drain field and thus creates a larger torque than the drain field (especially in the presence of dual gates), which makes its effect more pronounced. As the device sizes grow towards the micron regime, the increasing width (W) starts to overcome sticking, so the effect of the capacitive forces starts to overcome the effect of the Van Der Waals and dipolar forces.

\begin{figure}[t]
\centering
\includegraphics[width=\linewidth]{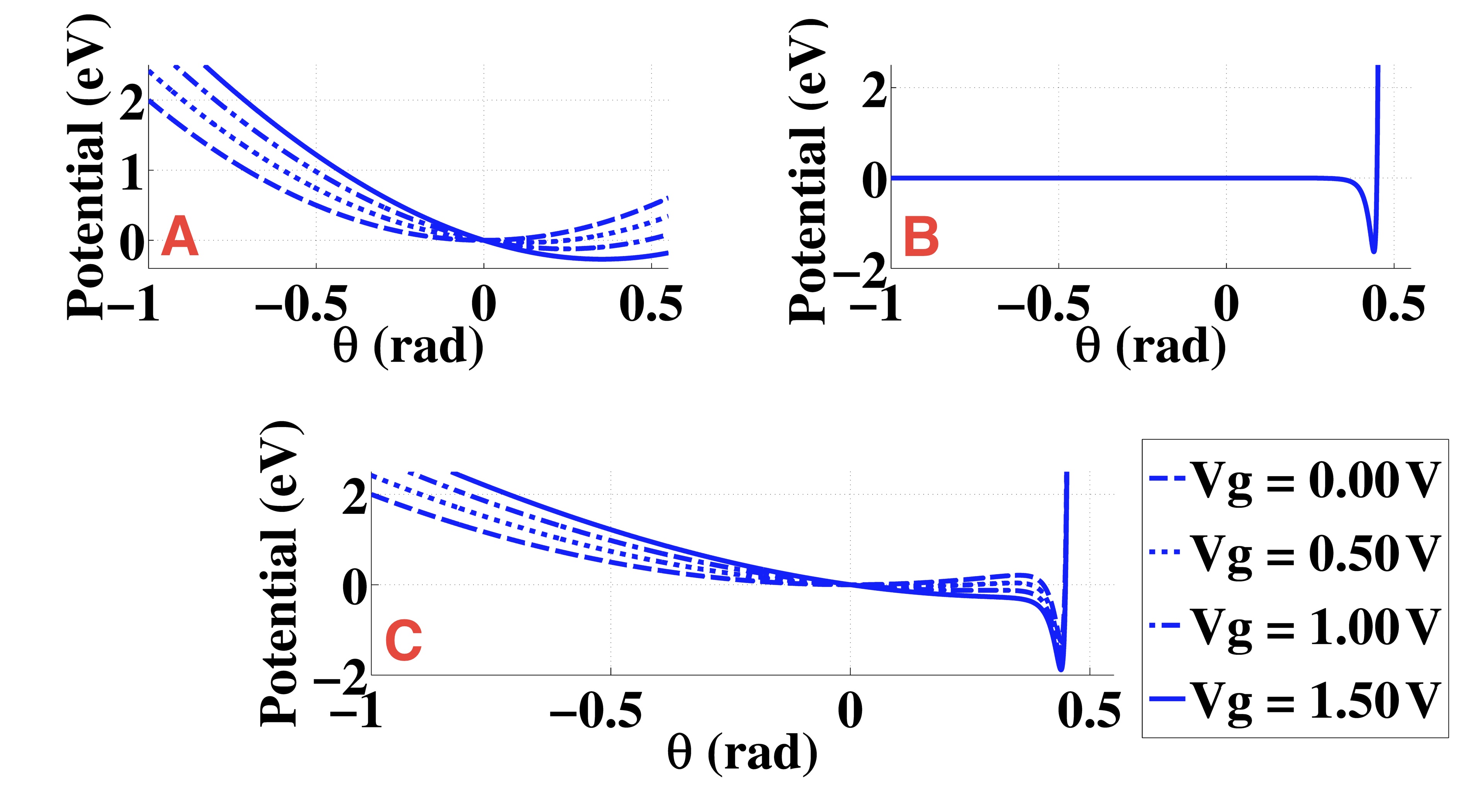}
\caption{{\bf (A)} The parabolic potential of the cantilever with one local minimum resulting from the combination of the stiffness and the external field terms. The local minimum shifts with the gate bias. {\bf (B)} The VDW potential resulting from the gate electrode with a stationary local minimum. {\bf (C)} A microscopic view of the parabolic potential varying with gate bias, in addition to a localized VDW potential from the contact or the capacitive destablization, creates a net profile that has one gate-dependent local minimum and one relatively fixed global minimum.}
\label{fig:profile}
\end{figure}

Fig. \ref{fig:profile}A shows the flexural parabolic potential, progressively shifted by the external electric field, causing a change in average angular orientation. Fig. \ref{fig:profile}B shows the separate Van der Waals potential that acts as a pull-in when the relay comes close enough to the drain. In combination of these two effects (Fig. \ref{fig:profile}C), there is a sweet spot where the potential profile between the metastable oscillator minimum and the stable Van der Waals minimum reaches a point of inflection, whereupon the cantilever angular coordinate makes an abrupt phase transition and locks onto the drain (in actuality this happens a bit earlier, once the hill separating the Van der Waals and parabolic minima reduces to around $\sim k_BT$). The pull-in has an important role to play. In its absence, the cantilever would be either too rigid to displace with an external field, or too floppy to resist thermal averages. In the absence of any other barrier, the thermal average over the conformational potential would tend to restore the system to its initial OFF state. This is a well known problem with a single well, analog computing - i.e., while it is energetically very pliable, it is highly corruptible with noise. In our geometry, we have an initial low energy analog variation combined with dipolar coupling, followed by a digital component at a later stage of the game, once the system abruptly falls into its ON state configuration in the well. In other words, it provides only a weak barrier to forward motion from the metastable to a stable state but a strong barrier for a return to origin. 

The barrier asymmetry is at the heart of the NEMFET operation, but relies on the intrinsic non-equilibrium nature of the dynamics. Left to itself long enough, the cantilever would always find a way to spontaneously jump out thermally from its metastable state to the stable state, i.e., it would bend over and stick to the drain given adequate time even in absence of a gate field assisting it. To define the transition between states, we will need to operate the voltage scanning faster than this spontaneous transition rate, so that the cantilever coordinate stays pinned to the metastable state until the voltage eliminates the barrier to the stable ground state. Our transport model must also probe this intermediate regime, and a simple steady state model will not capture the hysteresis associated with the metastable state. We discuss the details of this hysteresis in Section 5.

\begin{figure}[t]
\centering
\includegraphics[width=\linewidth]{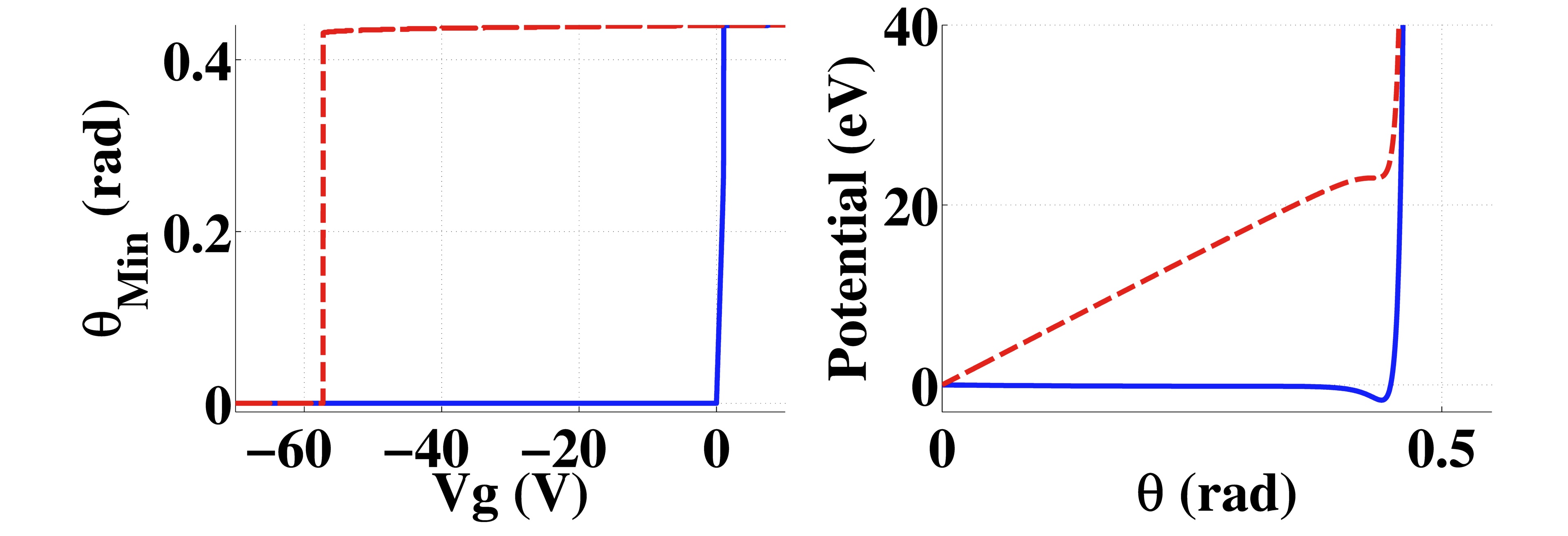}
\caption{Pull-in (shown by blue curves) and pull-out (shown by the red curves) mechanism rising from the NEMFET potential at positive and negative gate-biases. {\bf (Right)} The pull-in occurs when the positive gate-bias lowers the local maximum (barrier) between the two local minimums allowing the cantilever to move right. {\bf (Left)} The negative voltage shifts the local minimum from the VDW forces to higher energies, thus allowing the cantilever to shift left to the local minimum from the parabola.}
\label{fig:mechanism}
\end{figure}

Fig. \ref{fig:mechanism} shows the pull-in and pull-out dynamics mechanism arising out of the gate dependent alteration of the two well potential. The pull-in occurs when a sufficiently high positive gate bias is applied between the gate and the source electrode until at a critical voltage $V_{PI}$, the local maximum between the two minima disappears. The pull-out requires a high negative gate voltage $V_{PO}$ in order to break the Van der Waals adhesive bond. Described in terms of the potential landscape, the higher order flexural terms elevate the Van der Waals minimum until the latter becomes the metastable state and the cantilever configuration makes a reverse phase transition back to the horizontal coordinate.  The resulting potential and the average angle are shown in that figure. Clearly the $V_{PO}$ depends on the details of the Van der Waals potential and the higher order terms in the flexural modes, and can be reduced by adjusting the contact geometry, material and surface properties such as its roughness and quality. 

\section{Sharp switching and breaking the Landauer limit}
Let us try to work out the phase transition physics and the associated reduction in subthreshold swing. For a long stiff cantilever, the fluctuations about the mean can be ignored, so that $P(\theta) \propto e^{-\beta U(\theta)}$ becomes a delta function around the most probable angle $\theta^*$ set by the gate voltage $V_G$. We can expand around this extremum to give $U \approx U(\theta^*) + U^{\prime\prime}(\Delta\theta)^2/2$, (each $\prime$ denoting an angular derivative with $\theta$), so that the fluctuations in the Boltzmann distribution look like a Gaussian,
\begin{equation}
e^{\displaystyle -U(\theta)/k_BT} \approx e^{\displaystyle -U(\theta^*)/k_BT}e^{\displaystyle - U^{\prime\prime}(\Delta \theta)^2/2k_BT}
\label{gauss}
\end{equation}
with standard deviation $\sqrt{k_BT/U^{\prime\prime}}$. For a stiff cantilever compared to thermal fluctuations, we can approximate this Gaussian as a delta function around $\Delta\theta = 0$, which then integrated over angle gives us
\begin{equation}
\displaystyle I = \frac{2q}{h}\int_{\mu_1}^{\mu_2} dE M(E) \displaystyle\bar{T}\Bigl(E,\theta^*(V_G)\Bigr) \otimes F_T(E,V_G)
\end{equation}
evaluated at a single gate-dependent angle $\theta^*(V_G)$. Differentiating with the gate voltage, we get
\begin{equation}
\displaystyle\frac{\partial I}{\partial V_G} = \frac{2q}{h}\int_{\mu_1}^{\mu_2}
dEM(E)\Biggl[\frac{\partial\bar{T}}{\partial V_G} \otimes F_T + \bar{T}\otimes
\frac{\partial F_T}{\partial V_G}\Biggr]
\end{equation}
The tail of the thermal broadening function approaches the bandedge with an applied gate bias, making $\partial F_T/\partial V_G \approx \alpha_G \beta = \alpha_G q/k_BT$ for an enhancement mode transistor, while the  derivative of the first term in the bracket can be written as $(\partial\bar{T}/\partial V_G) = (\partial\bar{T}/\partial\theta^*)(\partial \theta^*/\partial V_G)$. The WKB tunneling term $\bar{T} = T_0\exp{[-2\kappa t_{air}(\theta^*)]}$ gives $\partial\bar{T}/\partial \theta^* = -2\kappa\bar{T}dt_{air}(\theta^*)/d\theta^*$. From geometry, $t_{air}(\theta^*) = h-L\sin{\theta^*} \approx L(\theta_0-\theta^*)$ for a long cantilever where $0 < \theta^* < \theta_0 \approx h/L \ll 1$. Then $dt_{air}(\theta^*)/d\theta^* \approx -L$, and we get the normalized transconductance as the separable sum of electronic and conformational
contributions
\begin{eqnarray}
\bar{g}_m &=& \displaystyle \frac{\partial \ln{I}}{\partial V_G} = \beta + 2\kappa L\Bigl(\frac{d\theta^*}{dV_G}\Bigr)\nonumber\\
&=& \bar{g}^{el}_m + \bar{g}^{conf}_m
\label{tradd}
\end{eqnarray}
assuming perfect gate control ($\alpha_G = 1$).  The normalized transconductance is inversely proportional to the subthreshold swing to within a constant factor of $\ln{10}/q$, so that the corresponding subthreshold swings add in parallel, reducing the overall value
\begin{equation}
S^{-1} = S_{el}^{-1} + S_{conf}^{-1}
\end{equation}
It is easy to see from Eq.~\ref{tradd} that the transconductance is already larger (and thus the subthreshold swing smaller) than the purely electrostatic limit $\beta$, and that this extra contribution can diverge if either the cantilever is very long ($\kappa L \gg 1$), or if the cantilever can be made to collapse abruptly with gate voltage $V_G$ so that $d\theta^*/dV_G \gg 1$.

\begin{figure}[b]
\centering
\includegraphics[width=\linewidth]{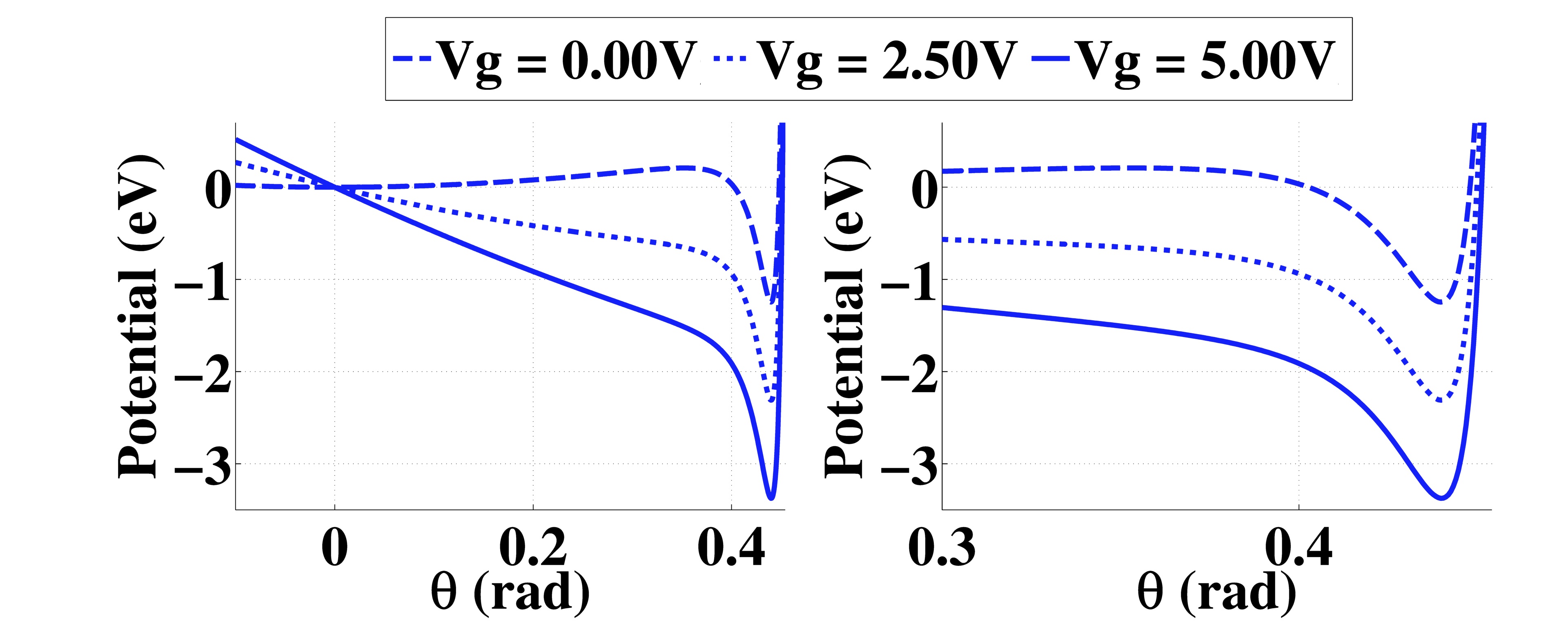}
\caption{The movement of the minimum, $\theta^*$, is shown in the potential profile. The $V_G$ shifts the metastable local minimum of the first well, until the local minimum disappears and jumps into the global minimum of the Van der Walls well.}
\label{fig:pot_stab}
\end{figure}

\begin{figure}[t]
\centering
\includegraphics[width=\linewidth]{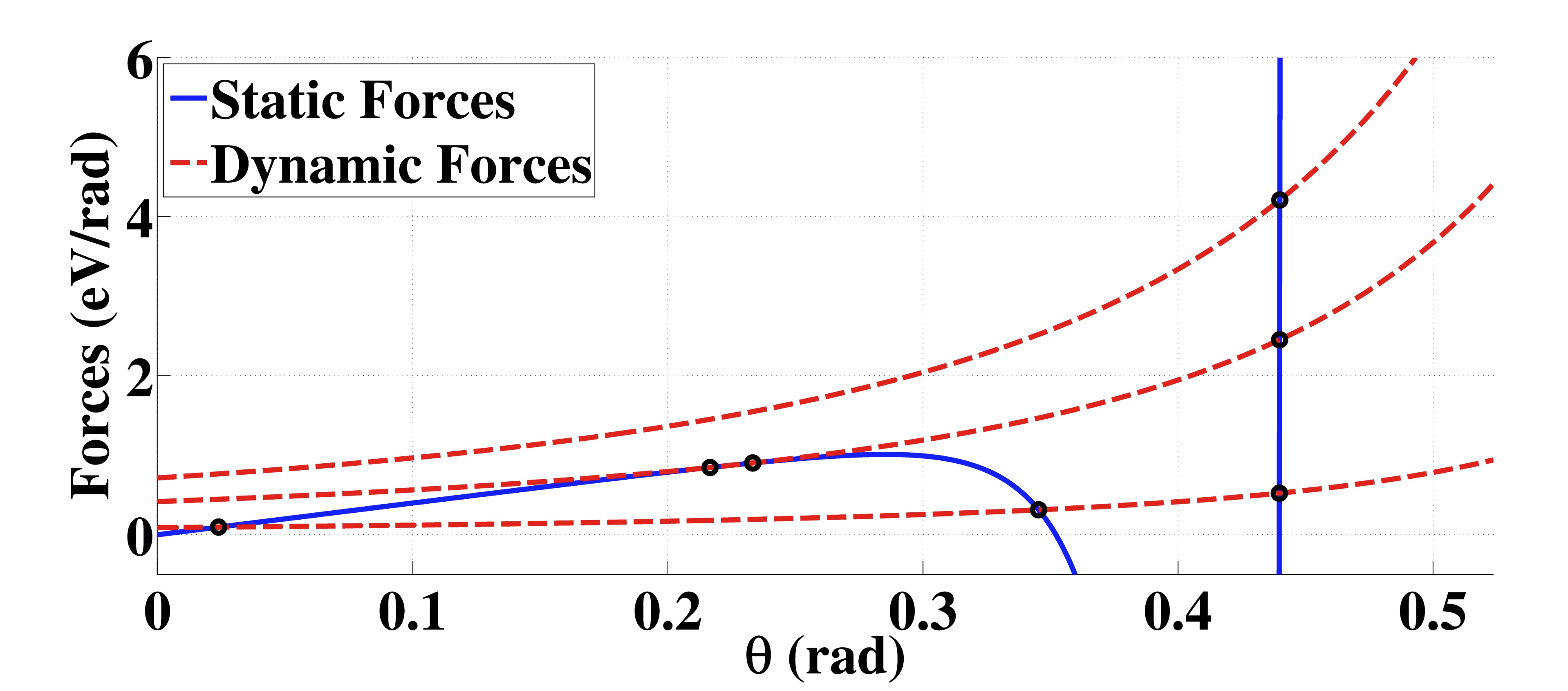}
\caption{The transcendental equation obtained by taking $dU(\theta^*)/d\theta^* = 0$ is shown  as the intersection of set of two curves. First set of curves (dashed red) are shifting vertically up as the $V_G$ value is increased. For the low values of the $V_G$, there are three in-section points corresponding to the three extrema: the local minimum from cantilever, the global minimum from Van der Waals, and the local maximum between two wells. At a the high enough $V_G$, local minimum and the local maximum collapse at a point of inflexion with $d^2U(\theta^*)/d(\theta^*)^2$, followed thereafter by only a single, global minimum. This movement of the minimums is the main reason of the phase transition in the relays.}
\label{fig:alpha_beta}
\end{figure}

The physics of the evolution of $\theta^*$ is easily seen from the potential profile in Fig. \ref{fig:pot_stab}. The gate shifts the minimum of the first well (the metastable state) in an analog format, till the cantilever coordinate falls into the Van der Waals well. The minimum can be obtained by taking $dU(\theta^*)/d\theta^* = 0$. This gives us a transcendental equation graphically shown in Fig. \ref{fig:alpha_beta} as the intersection of two sets of curves, one of which shifts vertically with increasing $V_G$. For low values of $V_G$, we have three intersection points corresponding to three extrema - the two well bottoms and the bump in between. At high enough $V_G$ however, the wells merge and there is only one point of intersection to the right, the global minimum at the bottom of the Van der Waals well, and this is what drives the phase transition. The transition occurs when the potential valley corresponding to the first metastable state reaches a point of inflexion, corresponding to the dual condition
\begin{eqnarray}
\displaystyle\frac{dU(\theta^*)}{d\theta^*} &=& 0 \nonumber\\
\displaystyle\frac{d^2U(\theta^*)}{d(\theta^*)^2} &=& 0
\end{eqnarray}

We can simplify the algebra considerably by replacing the Van der Waals potential with a stiff parabola around its minimum, and ignoring the oxide thickness for the moment. We can then write
\begin{eqnarray}
U(\theta) &\approx& \frac{1}{2}U_0^{Bend}\theta^2 + \frac{1}{2}k_{VDW}\Bigl(\theta-\theta_0\Bigr)^2 \nonumber\\
& - &\frac{\mu V_G\theta}{L(\theta_0-\theta)} -\frac{\epsilon_{air}A V_G^2}{2L(\theta_0-\theta)}
\end{eqnarray}
where $k_{VDW} = 57.14\epsilon L^2/\sigma^2$ around the Van Der Waals minimum. Setting its derivative equal to zero to extract the first metastable minimum, we get
\begin{equation}
(U_0^{Bend}+k_{VDW})\theta^*  -k_{VDW}\theta_0 =  
 \frac{2\mu V_G\theta_0 + \epsilon_{air}A V_G^2}{2L(\theta_0-\theta^*)^2} 
\end{equation}
which gives us the gate voltage dependence of the cantilever coordinate $\theta^*(V_G)$. Setting the second derivative to zero as well to get the inflexion condition gives as an additional equation, and from the pair of equations we can then extract the destablization point
\begin{equation}
(\theta^*)_{\rm{destab}}= \displaystyle \Biggl[\frac{U_0^{Bend} + 3k_{VDW}}{3(U_0^{Bend} + k_{VDW})}\Biggr]\theta_0
\end{equation}
We get a destabilization near $\theta \approx \theta_0/3$ when the bending forces dominate (this happens when we're far from the Van der Waals potential; the spring constant of the VDW term actually varies with angle if we take into account the nonlinearity that rapidly grows away from the sticking configuration). However, near $\theta \approx \theta_0$ to within a correction term $\Delta \theta \approx 2/\beta U^{\prime\prime}$ ignored in our Gaussian approximation, we will get an additional pull in as $k$ grows in strength. 

\begin{figure}[b]
\centering
\includegraphics[width=\linewidth]{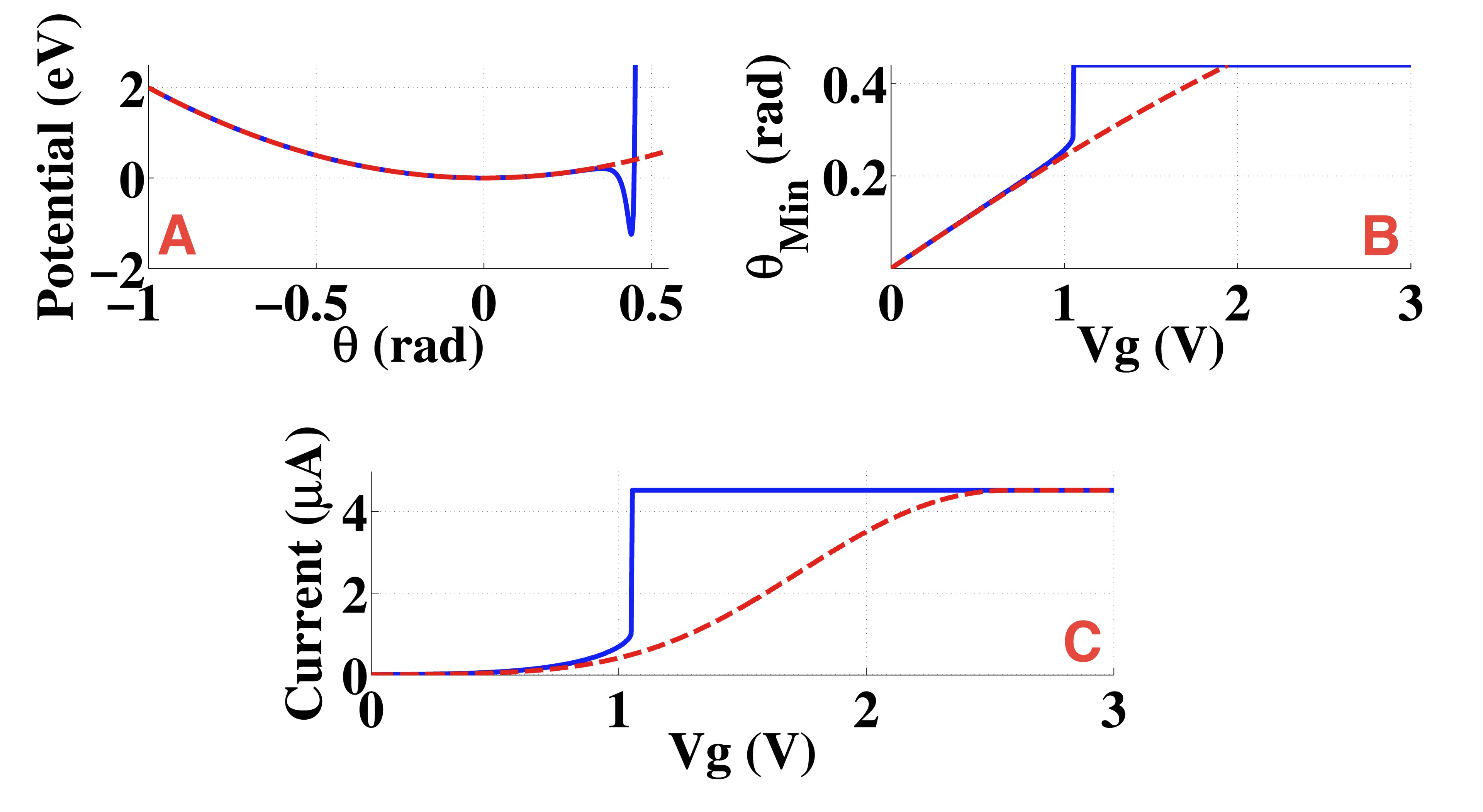}
\caption{{\bf (A} Comparison of the potential profile with the addition of VDW forces (shown by blue curves) and without the VDW forces (shown by red curves). {\bf (B} The angular movement of the cantilever with applied gate-bias with and without the VDW forces. {\bf (C)} With the VDW forces the cantilever gets pulled down instantaneously shown in red, causing an improved SS of $12 mV/dev$. Without the addition of VDW forces from the gate, the fastest the voltage swing is limited to thermal limit of 60 mV/dev as shown in dashed black line. Thus proving the importance of the VDW forces at the nano-scale limit on the sub-threshold swing.}
\label{fig:pullin}
\end{figure}

From the above equation, we can now extract the slope of the $\theta^*(V_G)$ curve, which enters the normalized transconductance $\bar{g}_m$
\begin{equation}
\bar{g}_m  =  \beta + 2\kappa \displaystyle\frac{\sqrt{\mu^2\theta_0^2 + U_{eff}\epsilon_{air} AL}}
{3U_{eff}(\theta^*-\theta^*_{\rm{destab}} )(\theta_0-\theta^*)}
\label{gm}
\end{equation}
where $U_{eff} = U_0^{Bend} + k_{VDW}$. 
As explained earlier, we can make the transconductance really high by either ramping up the total dipole moment $\mu$ or by collapsing at the destabilization point of the cantilever, $\theta^* \approx \theta^*_{destab}$. Near that point, the equation simplifies
\begin{equation}
\bar{g}_m\Bigr|_{Max} = \beta +  \kappa \displaystyle\frac{\sqrt{\mu^2\theta_0^2 + U_{eff}\epsilon_{air} AL}}{U_0^{Bend}\theta_0(\theta^*-\theta^*_{\rm{destab}})}
\label{eq:ss}
\end{equation}
The  second term can be interpreted as 
\begin{equation}
\bar{g}_m\Bigr|_{Max} =  \kappa L \displaystyle\frac{Q_{tot}}{U_0^{Bend}}\frac{\theta_0}{\theta^*-\theta^*_{\rm{destab}}}
\end{equation}
where $Q_{tot} = \partial U/\partial V_G$ is the total charge stored in the cantilever. The normalized transconductance reaches infinity at that point, which is an artifact of the fact that we only included quadratic terms in our potential, and ignored thermal fluctuations around the expected minimum $\theta^*$ assuming an infinitely stiff cantilever. As we discuss in the next section, including thermal fluctuations brings in a finite temperature term $\bar{g}^{conf}_m\propto \beta C$ averaged over all angles, which limits the subthreshold swing from its extreme zero value, although in practice this number can be (and has been measured to be) incredibly small. Eq.~\ref{gm} shows a second pull in near $\theta^* \approx \theta_0$ due to the Van Der Waals term, giving a second abrupt jump in transconductance. In Fig. \ref{fig:pullin}, we compare our results with and without the addition of VDW forces. In the absence of VDW, the performance is limited to 60 mV/dec and the cantilever angle $\theta$ increases slowly. With the inclusion of the VDW forces, the movement gets faster with the second potential minimum, lowering the SS to $12 mV/dev$. Once the gate voltage is large enough to remove the local maximum in Fig. \ref{fig:profile}, the switching occurs with the tip of the cantilever getting pulled in and adhering with the drain electrode.

\begin{figure} [ht!]
\centering
\includegraphics[width=\linewidth]{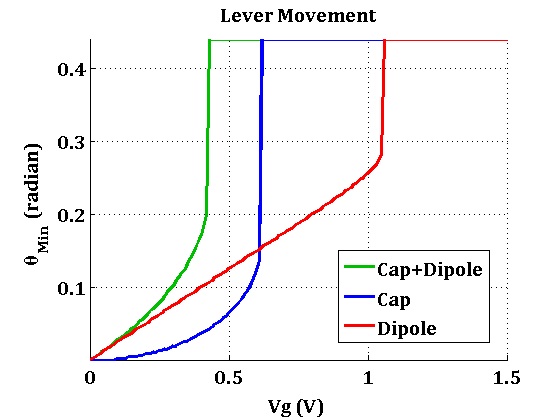}
\caption{Comparison of the model with the addition and removal of the dipole moments.. With the additional of dipole moments to the model the switching characteristics change. Without the dipole moment the pull-in happens at the 2/3 of the gap, where as with the strong dipole moments, the transition becomes much more linear instead of parabolic and the sticking occurs  at 1/3 of the gap. The model captures the exactly the same switching behavior as in \cite{5276844} when the dipole moment is not included and the same design parameters are used (shown as dashed.)}
\label{fig:dipole}
\end{figure}

As we scale the cantilever size, the gate to surface area reduces, making the capacitive forces weaker compared to the dipolar terms. For molecular cantilevers such as in sodium ion channels, the dipole moment dominates and is the main force behind the steep switching \cite{MacKinnon2003}. We can see this transition from electrostatic to dipole driven correlated switching in Fig.~\ref{fig:dipole}, where we scale the dipole term with the cantilever length. 

\section{Effects of finite temperature and finite scan rate}
So far, we calculated the subthreshold swing at room temperature, which means that we needed to include thermal fluctuations especially around the inflexion point. Such fluctuations will average out the sharp transition over an angular width $\Delta\theta \approx \sqrt{2k_BT/U^{\prime\prime}}$ that we had ignored earlier in Eq.~\ref{gauss}. Near the inflexion point however, this approximation breaks down. Starting with Eq.~\ref{landauer}, invoking an equilibrium thermal distribution $P$ for the configurational modes,  we can redo the calculation to extract the normalized transconductance 
\begin{equation}
(\bar{g}_m)_{conf} = \displaystyle\beta\Biggl[\frac{\bigl\langle \displaystyle\bar{T_0}\bigr\rangle\biggl\langle \displaystyle\frac{\partial U}{\partial V_G}\biggr\rangle -\biggl\langle \displaystyle\bar{T_0}\frac{\partial U}{\partial V_G}\biggr\rangle}{\displaystyle\bigl\langle\bar{T_0}\bigr\rangle}\Biggr]
\end{equation}
where the term in square brackets is the correlation term $C(\theta,V_G)$. 
For a long cantilever, $\partial U/\partial V_G$ computes to
\begin{equation}
\frac{\partial U}{\partial V_G} \approx -\frac{\mu \theta + \epsilon_{air}A V_G}{L(\theta_0-\theta)} 
\end{equation}
Note that while $\theta^*$ depended sensitively on $V_G$, we are now dealing with an independent variable $\theta$ that we will ultimately integrate over. The finite integration domain will keep the subthreshold swing finite, even near the inflexion point. However, the correlation term blows up at a particular angle, making the overall transconductance large but finite. The subthreshold swing has been measured to go down until as little as 0.1 mV/decade \cite{1609384}. 

After executing the voltage-gated phase transition, the cantilever sticks to the drain. It is the pull in from this adhesion that makes the cantilever angle change abruptly and causes its subthreshold swing to plummet. However, this means that for the reverse cycle, we will need to tear the cantilever away from the drain by catapulting it with a high negative gate bias out of the deep VDW potential back to the weaker parabolic potential. This adhesion creates a strong hysteresis loop, which bodes ill for subthreshold swing, because the {\it{average slope}} over the loop becomes quite large even though the walls of the loop remain steep.

\begin{figure} [t]
\centering
\includegraphics[width=\linewidth]{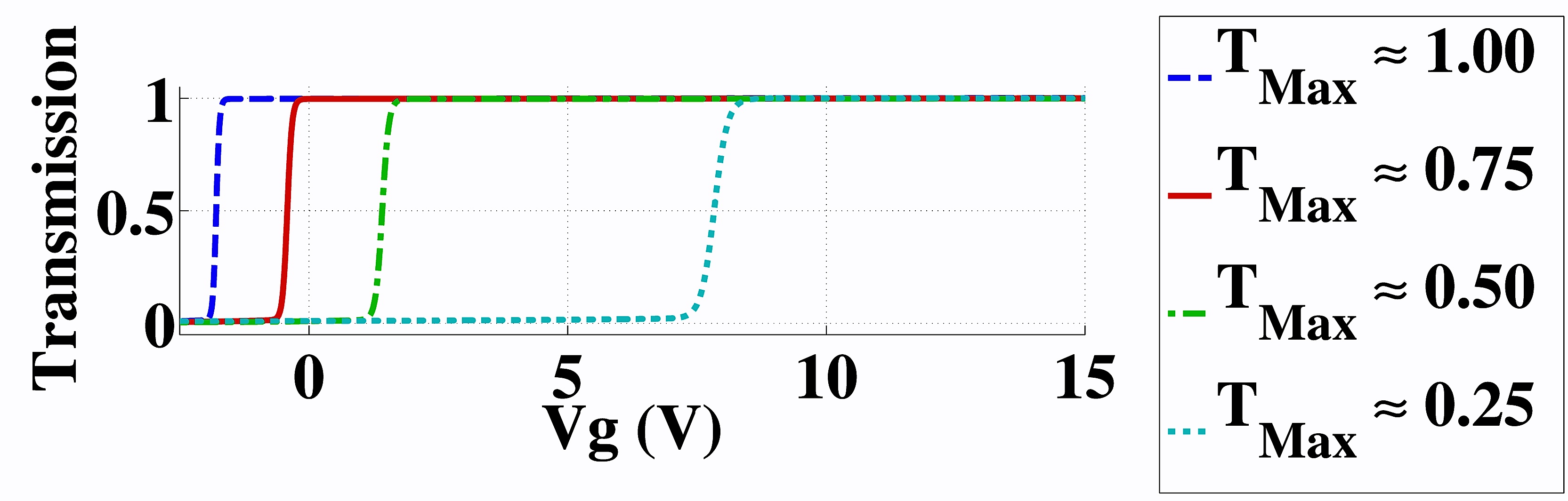}
\caption{The numerical solution of the Fokker-Plank equation coupled with the Landauer's theory results in switching voltages that are dependent on the scan-rate. As the scan-rate ($1/T_{max}$) increases, the switching voltages shifts to right, more positive values, causing a hysteresis. If the scan-rate is fast enough, the cantilever do not have time to follow the potential, resulting in a delayed version of the Boltzmann distribution. $T_{max}=1$ represents the case with no hysteresis where the scan-rate and the movement speed of the cantilever is comparable (Cantilever can follow the rate of change in the potential) (shown in {\bf A}). $T_{max}=0.25$ has a faster scan-rate than the cantilever can follow resulting in a delay in transition causing it to switch at higher gate bias. $T_{max}$ is measured in units of scan-time and it normalized to the case with no hysteresis.}
\label{fig:trans}
\end{figure}

Let us discuss how this hysteresis will arise out of our equations. The essential physics is governed by a voltage driven phase transition from a metastable to a stable state. For the forward swing, we move from a weak elastic potential to the Van Der Waals well, while for the reverse, we wait till the gate voltage pushes the Van Der Waals potential floor above the bottom of the elastic potential. It is clear that if the voltages varied very slowly near steady-state, i.e., given ample time, the cantilever would spontaneously stick by bending over and finding its global minimum, and thereafter the hysteresis would disappear. It is therefore essential to avoid a steady state calculation but invoke a finite scan rate $V_G(t)$, and do our calculations assuming that the time constants of the cantilever are slower than the scan rate of the gate voltage. While this is trivially true in most experiments, in our model, this means we can no longer assume an equilibrium Boltzmann distribution $P(\theta) \propto e^{-\beta\theta}$ but must solve the Fokker-Planck equation Eq.~\ref{eq4} with a time dependent gate voltage to extract the non-equilibrium behavior of the cantilever including its hysteresis. In Fig. \ref{fig:trans}, we show the distribution of electrons under a slow scan-rate, where the cantilever configuration tracks the local minimum of the potential profile, and also the fast scan-rate, where the conformational distribution function lags behind the evolution of the potential, generating the aforementioned hysteresis. For a given scan rate, we let our numerical Fokker Planck evolve until a maximum time set by the scan rate, and sample the configuration and the current at that intermediate time to plot the nonequilibrium distribution function. 

The microscopic model of hysteresis can be generalized for any system that have three different time scales: fast, slow, and intermediate. Hysteresis occurs when the transition occurs in the intermediate time scale between the fast and the slow. In the case of the NEMFETs, the slow time scale is the relaxation speed of the cantilever coordinate and the fast time scale is the maximum possible value of scan-rate of the top-gate (giving electrons almost no time to relax or drift with the potential profile). When the top-gate is driven with a slower scan-rate than the speed of the cantilever response function, the adiabatic approximation works. However as the scan-rate speeds up to an intermediate value, the switching gets postponed more and more, causing a scan-rate dependent switching voltage shown in Fig. \ref{fig:trans}. 

\section{Conclusions}
The hysteresis loop in the fabricated large devices (micron) and the smaller dimension biological and molecular devices (nanometer) vary and should not be ignored when scaling the devices to smaller sizes. Since the capacitive forces are weaker in the smaller devices, , the effect of the VDW dominates. As the devices get scaled aggressively to nanometer sizes, the nano-cantilevers will have wider hysteresis loops. This is where the dipole forces might help improve the performance of the devices, as shown in the Fig. \ref{fig:dipole}. Also the smaller the cantilever length, the harder it is for it to bend, so that the hysteresis width gets bigger. Means to reduce the hysteresis loop include scaling the width geometry of the cantilever to favor the dipole forces or underplaying VdW forces, designing a pull back electronic circuit with multigated structures, materials engineering to make the VdW well shallower, going to slower scan-rate or relying on ferroelectic phase transitions to liberate the cantilever from the Van Der Waals well \cite{6656832}. These effects however affect the subthreshold swing, so a proper optimization of the material and design landscape needs to be undertaken. The equations developed in this paper provide a starting point for such analysis.

\section{}

\bibliographystyle{IEEEtran}
\bibliography{IEEEabrv,NemFETs}

\end{document}